\documentclass[11pt]{article}
\usepackage[margin=3cm
]{geometry}

\usepackage{titling}
\usepackage{graphicx}% Include figure files
\usepackage{dcolumn}% Align table columns on decimal point
\usepackage{bm}% bold math
\usepackage{placeins}
\usepackage{authblk}
\usepackage{epstopdf}
\usepackage{amsmath}
\usepackage{braket}
\usepackage{setspace}
\usepackage[square,numbers]{natbib}
\usepackage{multibib}

\newcites{main,supp}{Main Text References ,Supplementary References}
\bibliographystylemain{unsrtnat}
\bibliographystylesupp{unsrtnat}

\newcommand{\fig}[5]{
	\begin{figure}
	\centerline{\includegraphics[width=#5\textwidth]{#1}}
\caption[#3]{\label{#2} \textbf{#3} #4}
\end{figure}}

\begin{document}

\title{Current-phase relations of few-mode InAs nanowire Josephson junctions}

\author[1,2]{Eric M. Spanton}
\author[3,4]{Mingtang Deng}
\author[3,5]{Saulius Vaitiek\.{e}nas}
\author[3]{Peter Krogstrup}
\author[3]{Jesper Nyg\aa rd}
\author[3]{Charles M. Marcus}
\author[1,2,6,*]{Kathryn A. Moler}

\affil[1]{Stanford Institute for Materials and Energy Sciences, SLAC National Accelerator Laboratory, Menlo Park, CA, USA}
\affil[2]{Department of Physics, Stanford University, Stanford, CA, USA}
\affil[3]{Center for Quantum Devices and Station Q Copenhagen, Niels Bohr Institute, University of Copenhagen, Copenhagen, Denmark}
\affil[4]{State Key Laboratory of High Performance Computing, NUDT, Changsha, 410073, China}
\affil[5]{Department of Physics, Freie Universit\"{a}t Berlin, Arnimallee 14, 14195 Berlin, Germany}
\affil[6]{Department of Applied Physics, Stanford University, Stanford, CA, USA}
\affil[*]{Corresponding Author: kmoler@stanford.edu}
\renewcommand\Authands{ and }

\date{\today}% 

\maketitle

\section{Introduction}

Gate-tunable semiconductor nanowires with superconducting leads have great potential for quantum computation \citemain{mourik, chang, Albrecht} and as model systems for mesoscopic Josephson junctions  \citemain{Beenakker,FurusakiPRB}. The supercurrent, $I$, versus the phase, $\phi$, across the junction is called the current-phase relation (CPR). It can reveal not only the amplitude of the critical current, but also the number of modes and their transmission. We measured the CPR of many individual InAs nanowire Josephson junctions, one junction at a time. Both the amplitude and shape of the CPR varied between junctions, with small critical currents and skewed CPRs indicating few-mode junctions with high transmissions. In a gate-tunable junction, we found that the CPR varied with gate voltage: Near the onset of supercurrent, we observed behavior consistent with resonant tunneling through a single, highly transmitting mode. The gate dependence is consistent with modeled subband structure that includes an effective tunneling barrier due to an abrupt change in the Fermi level at the boundary of the gate-tuned region. These measurements of skewed, tunable, few-mode CPRs are promising both for applications that require anharmonic junctions \citemain{Larsen,deLange} and for Majorana readout proposals \citemain{Hyart}.

\section{Main Text and Figures}

\begin{figure}[htb]
\centerline{\includegraphics[width=0.6\textwidth]{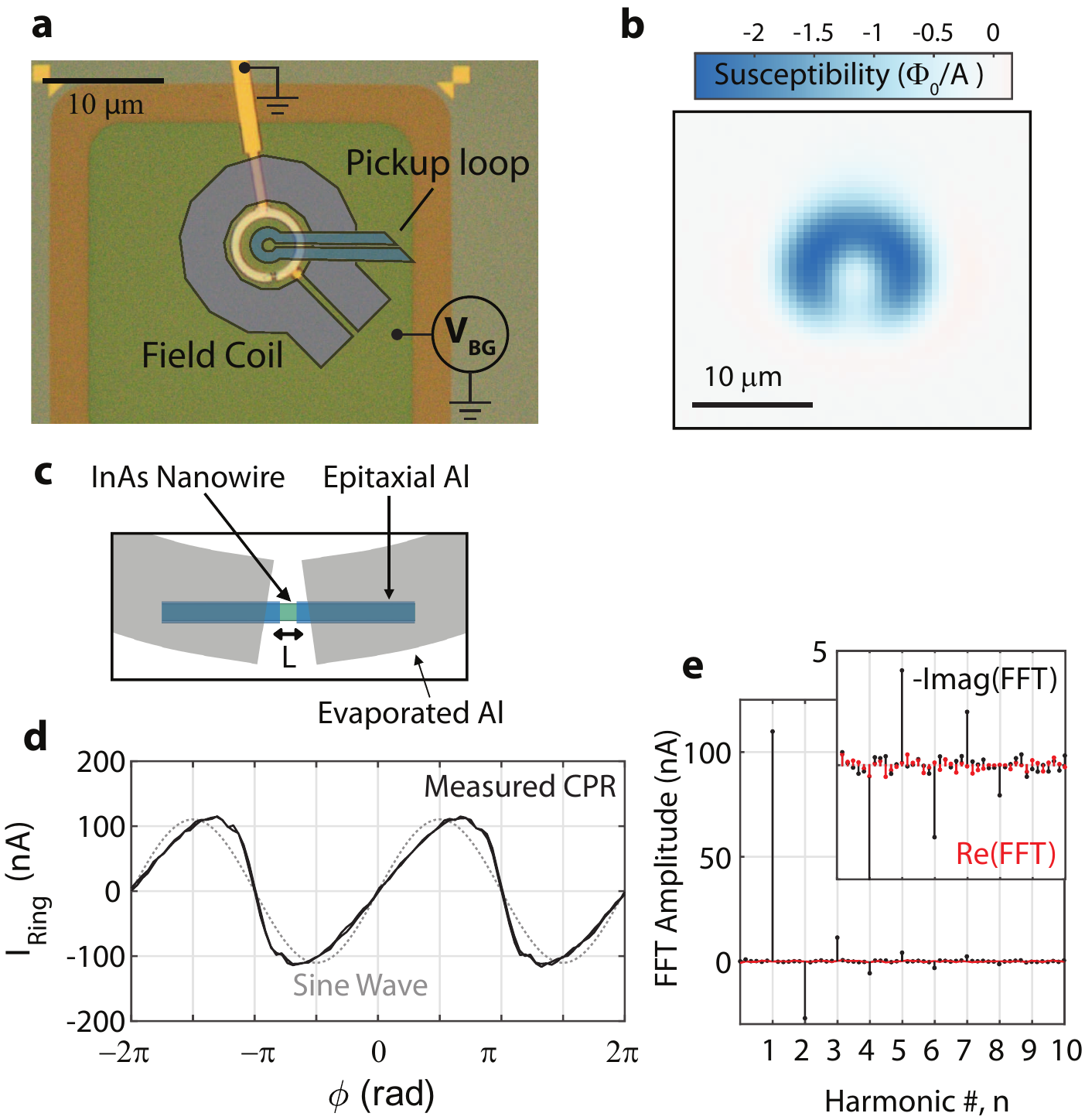}}
\caption[Current-phase relation of a few-mode InAs nanowire junction as measured by scanning SQUID microscopy shows significant forward skew.]{\label{fig:fig1} \textbf{Current-phase relation of a few-mode InAs nanowire junction as measured by scanning SQUID microscopy shows significant forward skew.} (a) Optical micrograph of an Al ring with a single epitaxial Al/InAs nanowire/Al junction. The size of the ring was matched to couple inductively to the geometry of the SQUID's pickup loop (blue; used to measure the current flowing around the ring) and field coil (purple; used to apply a phase difference across the nanowire junction). We used an Au bottom gate to tune the density of one nanowire in some of the measurements. (b) Local susceptibility image of a single-junction ring, showing the diamagnetic response of the evaporated Al ring. We used this signal to center the SQUID's pickup loop over the center of the ring. (c) Schematic of the epitaxial Al/nanowire/Al junction. The InAs nanowire (green) was coated on all sides with epitaxial Al (blue). We etched the Al of a small length (L) of the nanowire to form the normal junction region. Evaporated Al (gray) forms the ring. (d) An example of a measured CPR (black) of an $L$ = $150$ nm junction at $V_{BG}$ = 3.45 V and $T$ = 30 mK. The shape of the CPR is non-sinusoidal (a reference sine wave is presented in gray) and forward-skewed (the first maximum after $\phi =$ 0 is forward of $\pi / 2$). (e) Real and imaginary fast Fourier transform amplitudes (red and black, respectively) of the CPR taken from the interval $[-6\pi,6\pi]$. The imaginary component (black) shows that the shape of this CPR is well described by a sine Fourier series with 8 harmonics.}
\end{figure}

In superconductor-normal-superconductor (SNS) Josephson junctions, the critical current is predicted to be quantized as $I_C = N\Delta_0 / \hbar$ \citemain{Beenakker,FurusakiPRB}, where $N$ is the number of occupied subbands and $\Delta_0$ is the superconducting gap, as long as $N$ is sufficiently small and the junctions are short, ballistic, and adiabatically smooth. This quantization shows the number of Andreev bound states or modes, each coming from a single occupied subband, that carry the supercurrent. Single-mode junctions are desirable for the observation of Majoranas.  Quantized supercurrent has been difficult to realize experimentally, and it is difficult from transport measurements alone to determine which conditions for quantization are unmet.  In transport measurements of proximitized InAs and InSb nanowires, critical currents are far below the expected quantized values for perfectly-transmitting modes, and fluctuate with gate voltage \citemain{doh,Gunel,Abay,nilsson,liSciRep}. Nanowire Josephson junction qubits also display similar fluctuations in the measured resonant frequency \citemain{Larsen,deLange}. 
Here, we make direct, noncontact measurements (Fig.~\ref{fig:fig1}a) of Al/InAs nanowire/Al junctions (Fig.~\ref{fig:fig1}b,c) using an inductively-coupled \citemain{jackel} scanning SQUID (Methods, Ref. \citemain{sochnikov1, sochnikov2}) to measure the CPR. The contact between Al and  InAs nanowires is epitaxial \citemain{krogstrup}, greatly reducing the presence of in-gap states \citemain{chang}. In the low-temperature, few-mode limit, the shape of the CPR and its dependence on gate voltage, junction length, and temperature yield unique insights into the number and transmission of Andreev bound states.

The vast majority of CPRs we measured were forward-skewed; their first maximum after zero was forward of a sine wave (Fig.~\ref{fig:fig1}d). Fourier transforms of the CPRs revealed that they were well described by a Fourier series with up to seven harmonics (e.g. Fig.~\ref{fig:fig1}e). In the absence of time-reversal symmetry breaking, the theoretical CPR can be decomposed into a sine Fourier series \citemain{golubov}; fits to the experimental data include appropriate instrumental phase shifts (Methods, Supplementary Section 3).

The Fourier amplitudes ($A_n$) obtained by fitting the CPR characterize its shape. $A_1$ is the amplitude of the 2$\pi$-periodic component and is approximately $I_C$. The `shape parameters' $a_n \equiv (-1)^{n+1} A_n/A_1$ characterize the shape of the CPR independent of its amplitude. Positive $ a_2$ indicates a forward-skewed CPR; negative $a_2$ indicates a backward-skewed CPR. Higher-order terms yield more detailed information about the shape of the CPR and can be used to extract information about multiple modes, to compare to CPR theories, and to differentiate between the effects of elevated temperature and lowered transmission.

The amplitude and shape of the CPR of an $L$ = 150 nm nanowire junction fluctuated as a function of applied bottom gate voltage, $V_{BG}$. The fluctuations were most dramatic close to depletion of the nanowire as seen in Fig.~\ref{fig:fig2}. The most forward-skewed CPR observed in this study and a backward-skewed CPR both occurred in the same gate-tuned junction (Fig.~\ref{fig:fig2}a). Fluctuations in $A_1$ remained similar in amplitude for all $V_{BG}$ (Fig.~\ref{fig:fig2}b), while fluctuations in $a_n$ decreased with increased $V_{BG}$ (Fig.~\ref{fig:fig2}c) and with elevated temperature (Supplementary Section 1).

$A_n$ and $a_n$ both displayed peaked behavior near depletion, where we expect a single subband to be occupied (Fig.~\ref{fig:fig3}a,b). The peaks in both $a_2$ and $A_n$ were asymmetric, with the side at more positive $V_{BG}$ appearing more broad (Fig.~\ref{fig:fig3}a,b). The most forward-skewed CPR occurred in this peaked regime (Fig.~\ref{fig:fig3}c).

We observed a backward-skewed CPR only for a narrow gate voltage range in the gated junction, and never in ungated junctions (Fig.~\ref{fig:fig4}a). $a_2$ was negative in the backward-skewed region (Fig.~\ref{fig:fig4}c) and was nearly coincident with a minimum in $A_1$ (Fig.~\ref{fig:fig4}b). $a_3$ switched sign when $a_2$ was at its most negative. Our observation of a backward-skewed CPR cannot be explained by noise rounding, as in Ref. \citemain{English}, as the junction studied here was not near the hysteretic regime (Supplementary Section 4). 

Junctions with different lengths (Fig.~\ref{fig:fig5}) exhibited a range of $I_C$ and forward-skewness (Fig.~\ref{fig:fig5}a). $A_1$ did not depend strongly on the length of the junction ($L$) (Fig.~\ref{fig:fig5}b) and $a_2$ trended downward with increasing $L$, albeit with significant scatter (Fig.~\ref{fig:fig5}c). We detected no clear correlation between $A_1$ and $a_2$ (Supplementary Section 6), which suggests that variations in the Fermi level from junction to junction were responsible for fluctuations in the amplitude and shape of the CPR.

For SNS junctions in the short-junction limit (when $L$ is much smaller than the superconductor's coherence length), each occupied subband (mode) leads to a single Andreev bound state whose properties depend only on its normal-state transmission ($\tau$) \citemain{Beenakker}. The CPR is predicted to be sinusoidal when $T \sim T_C$ or when $\tau\ll 1$, but forward-skewed at low temperatures and high transmissions, as evident in the short-junction expression for the CPR:

\begin{equation}\label{eq:SJ_theory}
I(\phi) = \frac{e \Delta_0(T)}{2 \hbar} \sum_{p=1}^{N} \frac{\tau_p sin(\phi)}{[1-\tau_p sin^2(\phi/2)]^{1/2}} tanh(\frac{\Delta_0(T)}{2 k_B T} [1-\tau_p sin^2(\phi/2)]^{1/2}), 
\end{equation} where $T$ is the temperature, $k_B$ is the Boltzmann constant, $\hbar$ is the reduced Planck constant, $\Delta_0(T)$ is the superconducting gap, and $e$ is the electron charge. 

To estimate the number of modes ($N$) and transmission ($\tau$) of the junctions, we first assumed $\tau$ was the same for all modes and $T$ based on our mixing chamber temperature. For $V_{BG} > 4$ V in Fig.~2, $a_2 \approx 0.2$ and $A_1 \approx 100$ nA; we estimate that $\tau \approx 0.8$ and $N \approx 4-5$. For the CPRs measured on many junctions in Fig.~5, $N$ varies from $0-10$ and $\tau$ varies from 0.5-0.9 for CPRs with $A_1 > 10$ nA. Therefore, the junctions we studied are in the few-mode regime and often have very high transmission.
Peaks in $A_n$ versus gate voltage at low densities indicate resonant tunneling behavior (Figs.~3 and 4.) Eq.~\ref{eq:SJ_theory} remains valid for a junction with tunnel barriers, with $\tau$ given by the Breit-Wigner transmission \citemain{Beenakker}:

\begin{equation}\label{eq:BW}
\tau = \frac{\Gamma_L \Gamma_R}{(E_F-E_0)^2+1/4(\Gamma_L+\Gamma_R)^2}, 
\end{equation} where $\Gamma_{L,R}$ are tunnel rates into the left and right barriers, respectively, and $E_0$ is the energy of the resonance with respect to the Fermi energy, $E_F$. Eq.~\ref{eq:BW} is valid for SINIS junctions only when $\Gamma_L + \Gamma_R \gg \Delta_0$ and shows that perfect transmission ($\tau =1$) requires the left and right tunnel barriers to have identical tunnel rates.
Fig.~3c fits well to Eq.~\ref{eq:SJ_theory} with four parameters: $T$, $\tau$, a scaling parameter ($\epsilon$), and a phase shift ($\phi_0$). Fixing $T$ to its measured value, 0.03 K, gives a good fit with $\epsilon$ and $\phi_0$ consistent with our experiment, with a best-fit $\tau = 0.98$. Allowing $T$ to vary gives a best-fit value of $T$ = 0.13 K and $\tau = 1.00$ (Supplementary Section 2). These results indicate perfect or nearly perfect transmission.

We attribute the high-transmission mode (Fig.~\ref{fig:fig3}c) to tunneling into a resonant mode with symmetric barriers. Perfectly symmetric tunnel barriers are unlikely to accidentally occur due to disorder-induced quantum dot behavior, so their origin is more likely an intrinsic barrier. Symmetric barriers could arise at the InAs/Al interface and/or at the border between etched and Al-coated parts of the nanowire (Fig.~\ref{fig:fig1}c) due to different band bending in those regions resulting in momentum mismatch. The effects of mismatch can be approximated as a delta-function potential at the interface \citemain{FurusakiPRB}.

Two types of simulations that include the effects of wave function mismatch qualitatively matched the behavior displayed in Figs.~2 and 3. Simulations in accordance with Ref. \citemain{FurusakiPRB} revealed peaked behavior at low gate voltage, and in particular reproduced the presence of a highly forward-skewed mode and the shape and asymmetry of peaks in the fitted CPR parameters (Fig.~\ref{fig:fig3}a,b). They also reproduced a reduction of fluctuations $a_n$ with higher electron densities. Tight-binding simulations that accounted for the cylindrical shape of the nanowire geometry also qualitatively matched our observations, including the presence of peaks with a spread of $a_2$ at low densities (Supplementary Section 5) \citemain{kwant}.

The backward-skewed CPR (Fig.~\ref{fig:fig4}), however, does not arise in theories of short-junction SNS or resonant tunneling behavior. Backward-skew can theoretically arise from pair breaking in the superconducting leads, which is unlikely here due to the low $I_C$ of the junction \citemain{golubov}. The origin of a gate-voltage-dependent backward-skewed CPR remains an open question. However, based on the sharp gate dependence and the minimum in $I_C$ coincident with the backward-skewed CPR, we speculate that a strongly gate-dependent tunnel barrier may lead to effective pair-breaking in the proximitized nanowire and therefore the backward-skewed CPR.

The shape of the CPR has important consequences for nanowire Josephson junction devices. Our observation of a single $\tau \sim 1$ mode that occurs at a peak in $I_c$, rather than a step, suggests that weak, symmetric tunnel barriers can form due to wave function mismatch in the nanowire. The resonance condition (Eq.~\ref{eq:BW}) leads to the maximum possible tunability of the skewness, and therefore the anharmonicity of the Josephson potential, with small gate voltages. Control of  anharmonicity can tune nanowire-based qubits to a flux-qubit-like regime near $\phi = \pi/2$ \citemain{deLange}.

The skewness of the CPR is also important in determining the behavior of the junction in an applied field (the Fraunhofer diffraction pattern). Although such measurements are difficult to achieve on nanowire junctions, our data indicate that the shape of the CPR can vary substantially (and even become backward-skewed) for high-quality nanowire junctions, which should lead to noticeable effects in the Fraunhofer pattern. Therefore, the likelihood of a non-sinusoidal CPR must be considered when attempting to extract the spatial distribution of current in Josephson junctions \citemain{Hart,Allen}.

Inductive measurements of nanowire Josephson junctions with spin-orbit coupling \citemain{Alicea} or quantum spin Hall junctions \citemain{FuKane} have been proposed as a readout mechanism for the parity of Majoranas. We have measured the CPR with the requisite sensitivity for non-contact readout of Majorana parity and observed the CPR of a single occupied subband that has close-to-perfect transmission. Challenges remain in realizing Majorana readout. First, the in-plane fields required to create Majoranas in InAs nanowires are a major technical challenge for scanning SQUID microscopy. Second, the time scale of the measurements presented here (up to tens of minutes) may not allow a measurement of a full CPR before environmental factors cause the parity to switch. Nevertheless, our sensitivity to such parity switching would allow us to use noise measurements to observe such environmental switching as long as the parity lifetime is longer than  $\approx10$ $\mu s$.

\fig{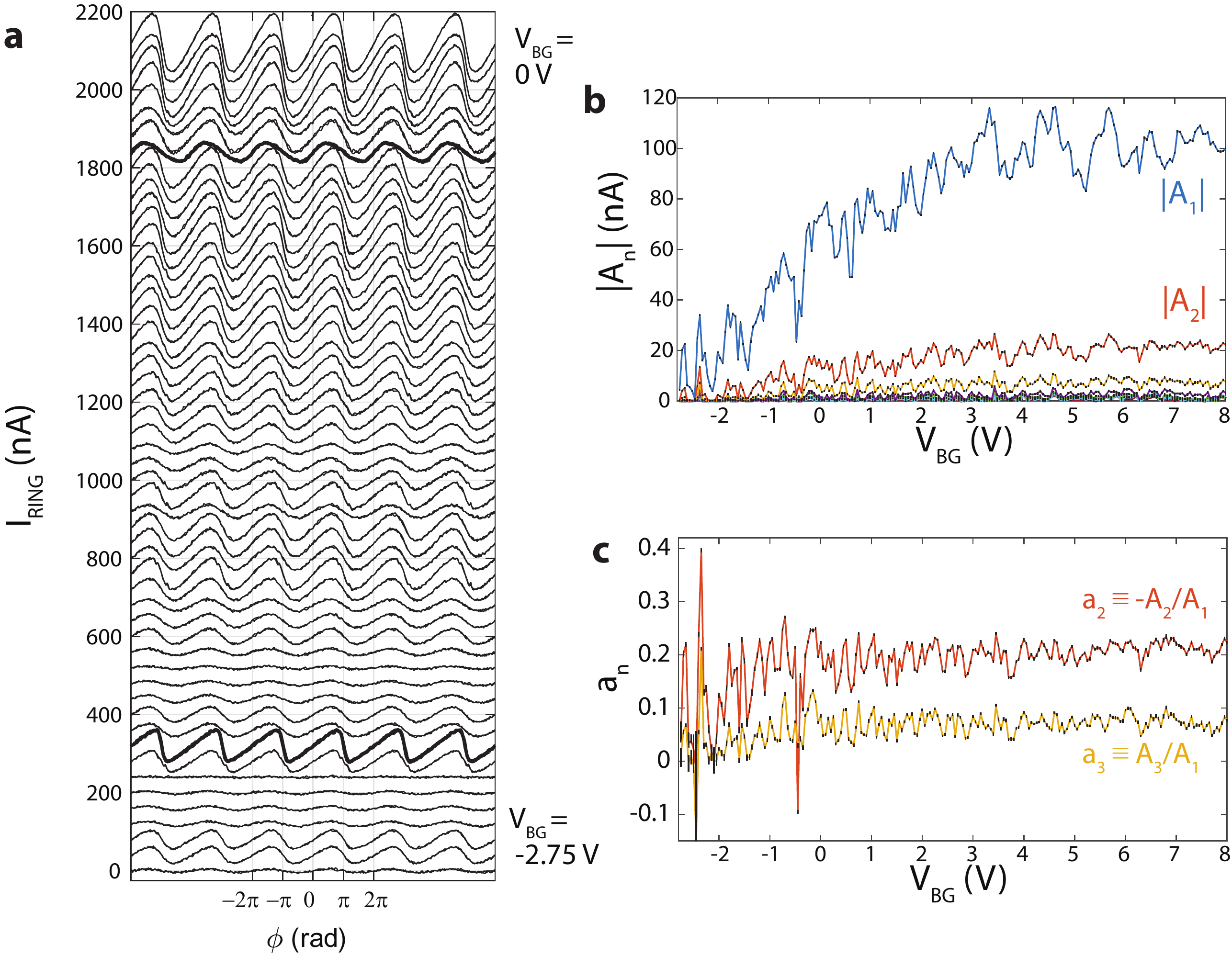}{fig:fig2}
{Fluctuations in the forward skew and amplitude of the current-phase relation of a bottom-gated junction with gate voltage}
{(a) CPRs (offset for clarity) versus bottom gate voltage ($V_{BG}$) for an $L$ = 150 nm single-junction ring. $V_{BG}$ = -2.75 V is close to depletion of the nanowire. The amplitude and shape of the CPR fluctuated strongly with gate voltage, particularly at the low gate voltages presented here. The most forward-skewed CPR and a backward-skewed CPR both occurred at low gate voltages (bolded curves). (b) Absolute value of the fitted harmonic amplitudes ($A_n$) versus $V_{BG}$. Non-zero harmonics for up to n = 7 were observed, and fluctuations were present for all of them. (c) Fitted shape parameters ($a_n \equiv (-1)^{(n+1)} A_n/A_1$) versus $V_{BG}$. $a_2$ was primarily positive, indicating a forward-skewed CPR.}{0.9}

\fig{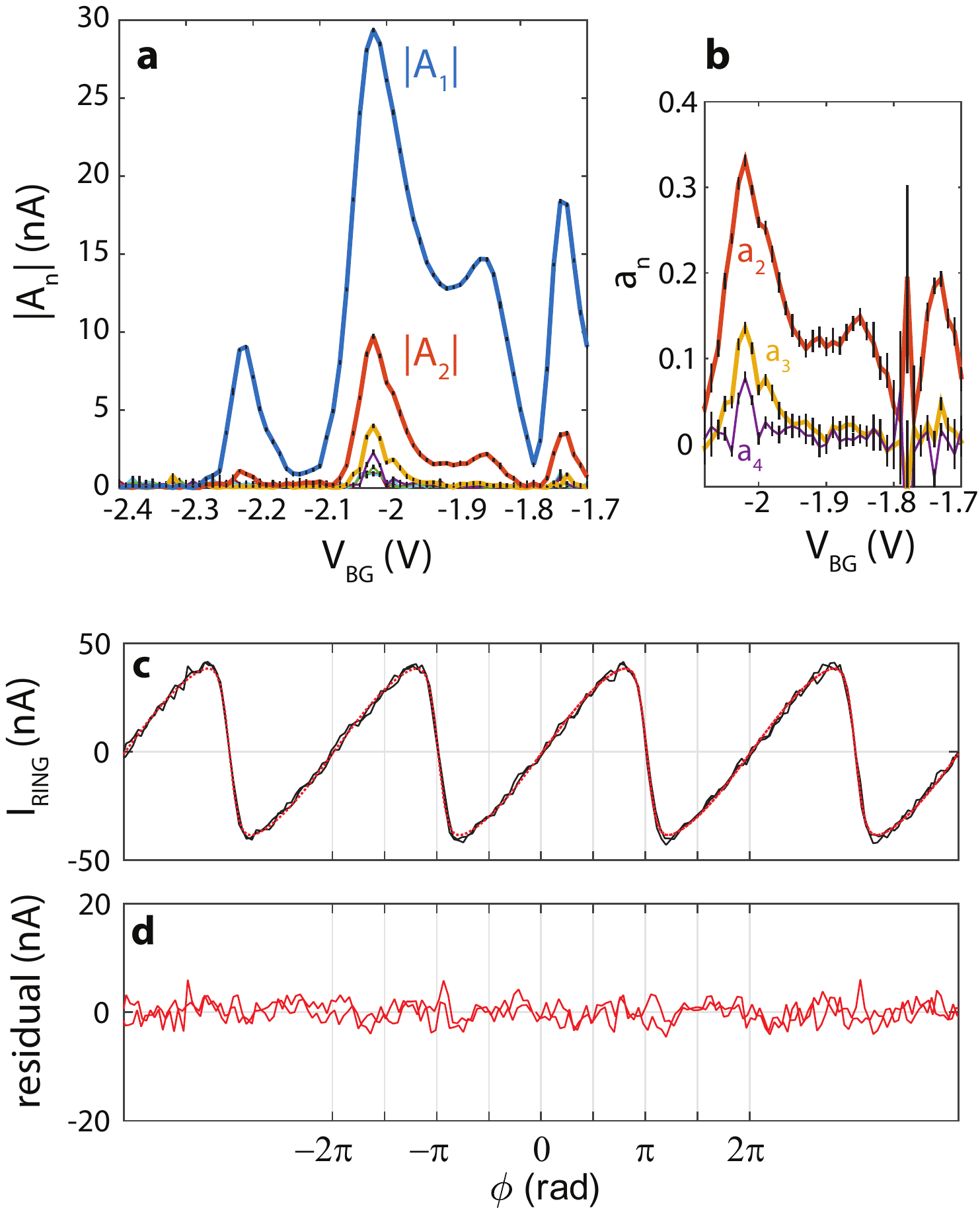}{fig:fig3}
{Peak-like behavior in the shape and skewness of the CPR at low gate voltages.}
{(a,b) Harmonic amplitudes ($|A_n|$) and shape parameters ($a_n$) versus $V_{BG}$, respectively, close to depletion with smaller gate voltage steps than Fig.~2. The shape of peaks in $a_2$ are cusp-like and asymmetric, consistent with simulations of a junction with wave function mismatch which leads to resonant tunneling (Supplementary Section 5). The relative peak positions and heights in (a) and (b) are similar to those in Fig.~2 with an overall shift in $V_{BG}$, indicating that the features are robust to large sweeps in $V_{BG}$. (c) Measured CPR (black) at $V_{BG}$ = -2.35 V, from the gate sweep in Fig.~2. This CPR is the most forward-skewed CPR we observed and occurred in the low density, peak-like regime. A free temperature fit, which allows for elevated electron temperature, yielded the following fitted parameters: $T$ = 0.13 K, transmission $\tau =$ 1.00, and a scaling factor that accounts for slight errors in the positioning of the SQUID $1+\epsilon$ = 1.08. A fit with fixed $T$ = 30 mK and free $\epsilon$ gives $\tau =$ 1.00. (d) Residuals of the short junction fit presented in (c). The fit is of good quality with no obvious structure in the residuals.}{0.7}

\fig{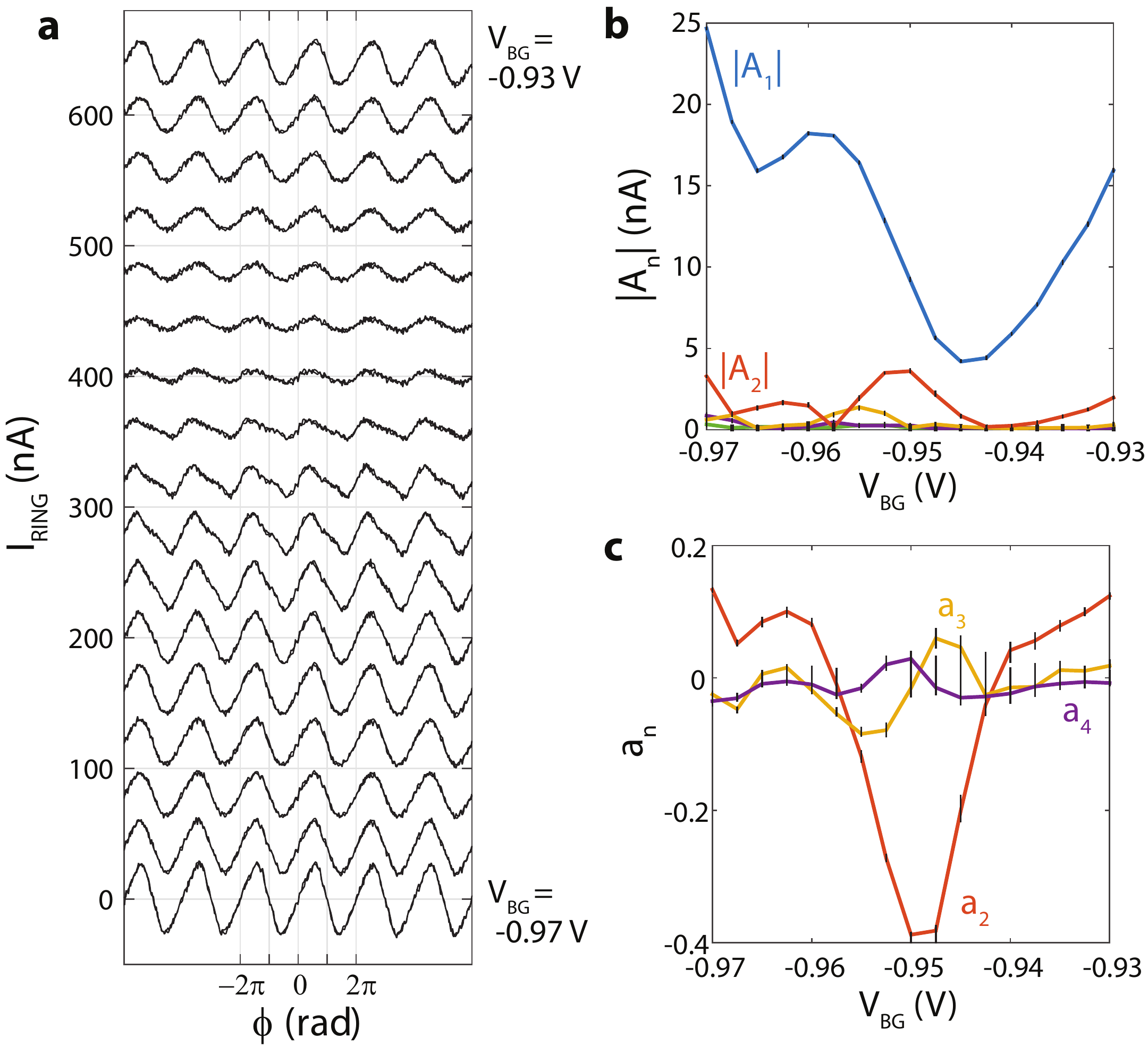}{fig:fig4}
{Backward-skewed current-phase relations for a narrow range of gate voltages.}
{(a) CPRs versus $V_{BG}$ with offsets for clarity, close to the region with backward-skewed CPRs (with a first maximum before $\phi = \pi/2$). (b,c) Harmonic amplitudes ($|A_n|$) and shape parameters ($a_n$) versus $V_{BG}$, respectively. The first maximum of the shape precedes $\phi  = \pi/2$ for a narrow range of gate voltages (a) and correspondingly $a_2$ becomes negative (c). The maximum backward-skewness is associated, but not completely coincident, with a local minimum in the amplitude, $A_1$ (b,c).}{0.9}

\fig{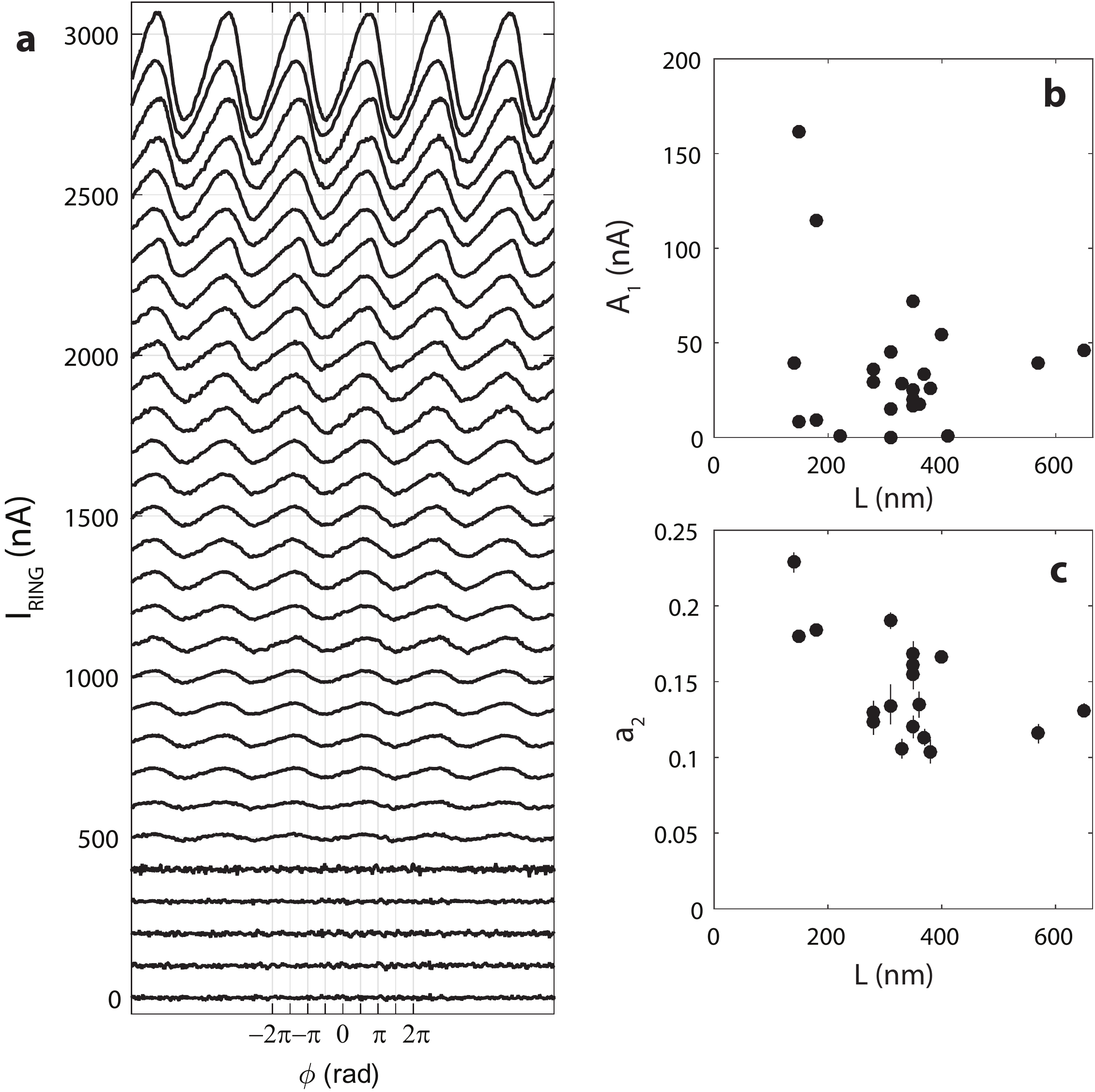}{fig:fig5}
{The forward-skewness of the current-phase relation for many junctions with various lengths.}
{(a) CPRs of many InAs nanowire junctions with various lengths, at low temperatures ($T < 50$ mK) and with no applied gate voltage. The CPRs are ordered by their overall amplitude. The shape and amplitude of the CPR varied substantially, and the degree of forward-skewness was not related to the amplitude. (b) Fitted amplitude of the first harmonic ($A_1$) versus the nominal etched length of the junction ($L$). $A_1$ is approximately the critical current ($I_C$). $A_1$ does not show an obvious trend with length, therefore junction-to-junction variation (presumably of the chemical potential) is more important than L in determining $A_1$. (c) Fitted shape parameter ($a_2 \equiv –A_2/A_1$), which is directly proportional to the forward-skewness of the CPR, versus $L$. The forward-skewness weakly trended downward with longer junctions, indicating that the transmission of the junction was decreasing. We also detected large variations in forward-skewness for junctions of the same length. Error bars are 90$\%$ confidence intervals obtained from bootstrapping (see Methods).}{0.9}

\FloatBarrier

\section{Methods}
InAs nanowires were grown along the $[0001]_B$ direction and Al was deposited epitaxially to fully coat the nanowires \citemain{krogstrup}. A length, $L$, of the coated nanowire was etched to form a Josephson junction. We evaporated Ti/Al (5 nm/120 nm) ex situ to form the ring geometry required for CPR measurements. We measured the dimensions of the nanowires using scanning electron microscopy after the scanning SQUID measurements.

We measured three samples, each with many single-junction rings, in a dilution refrigerator. In the gated junction, a gold bottom gate made of Ti/Au (5nm/20 nm) with a AlO$_x$ dielectric (40 nm) was used to tune the density of the nanowire. A number of devices were hysteretic; scanning electron microscopy revealed that they were under-etched (inset of Supplementary Fig.~S5).
We used a SQUID microscope in a dilution refrigerator with a nominal base temperature of $\sim 30$ mK \citemain{gardner,huber}. Temperatures were measured at the mixing chamber plate using a Rox thermometer. Data in the main text were all taken at $T<50$ mK.
We centered the scanning SQUID's pickup loop and field coil over the ring using the diamagnetic response of the evaporated Al to navigate (Fig.~1b). With the SQUID centered over the ring, we swept the current through our local field coil ($I_{FC}$) sinusoidally at $\approx 200 $Hz and recorded the flux through the SQUID’s pickup loop ($\Phi_{PU}$) (Fig.~\ref{fig:fig1}a). We set the amplitude of $I_{FC}$ to thread multiple flux quanta through the ring. We subtracted linear and experimental background from the measured $\Phi_{PU}$ versus $I_{FC}$ to account for the diamagnetic response of the ring and imperfect geometric cancellation of the applied local field.

In single-junction rings (Fig.~\ref{fig:fig1}a), the CPR can be directly measured inductively as long as the self inductance of the ring is small, $L_{self} < \Phi_0 / 2 \pi I_c$, where $\Phi_0$ is the superconducting flux quantum \citemain{jackel}.
We converted measured $\Phi_{PU}$ versus $I_{FC}$ to $I$ versus $\phi$ (the CPR) using the periodicity of the signal and calculations of the mutual and self inductances \citemain{brandt}. Although we included corrections for self-inductance effects in the conversion, they were not important in determining the shape of the CPR because the (non-shorted) junctions were always in the limit of $\beta = 2 \pi L_{self} I_C / \Phi_0 \ll 1$. We confirmed our calculation of the self-inductance by measuring the height of steps in the response of hysteretic rings (Supplementary Section 4). Errors in centering the SQUID's pickup loop over the ring of a few microns would result in a systematic error of $\approx 10\%$ in the extracted current and Fourier components $A_n$. The shape parameters $a_n$, however, are insensitive to this form of systematic error.
We fitted both forward and backward sweeps of the CPR to extract their Fourier components. We fit the CPRs presented in Figs.~2-4 using:

\begin{equation}\label{eq:fixed_fits}
I(\phi) = \begin{cases} \sum_{n=1}^N A_n sin (n (\phi + \phi_{FW})) &\mbox{if }d\phi/dt > 0 \\ 
\sum_{n=1}^N A_n sin (n (\phi + \phi_{BW})) &\mbox{if } d\phi/dt < 0 \\ 
\end{cases}, 
\end{equation} which fully accounted for the shape of the CPR. The phase shifts between forward and backward sweeps ($\phi_{bw,fw}$) were subtracted from $\phi$ in figures presented in the main text.
In Fig.~\ref{fig:fig5}, some CPRs exhibited phase shifts between harmonics. The harmonics were shifted in opposite directions for forward and backward sweeps, indicating that they were instrumental rather than intrinsic to the junction (see Supplements). Phase shifts between harmonics were not present in the CPRs in Fig.~2-4, as is evident in the lack of an out-of-phase component in the fast Fourier transform (Fig.~\ref{fig:fig1}e).

\FloatBarrier

\section{Acknowledgments}
We thank Sean Hart, John Kirtley and Carlo Beenakker for useful discussions and Christopher Watson, Zheng Cui, and Ilya Sochnikov for useful discussions and experimental assistance.

The scanning SQUID measurements were supported by the Department of Energy, Office of Basic Energy Sciences, Division of Materials Sciences and Engineering, under Contract No. DE-AC02-76SF00515. Nanowire growth and device fabrication was supported  by  Microsoft  Project  Q, the Danish National Research Foundation, the Lundbeck Foundation,  the  Carlsberg Foundation, and the European Commission. C.M.M. acknowledges support from the
Villum Foundation.

\section{Author Contributions}
P.K. and J.N. developed the nanowire materials, M.D. and S.V. fabricated the devices and E.M.S. performed the scanning SQUID measurements, analyzed the data, and performed simulations. E.M.S. and K.A.M. wrote the manuscript with input from all coauthors.

\bibliographymain{refs}

\renewcommand{\thefigure}{S\arabic{figure}}
\renewcommand{\theequation}{S.\arabic{equation}}
\renewcommand{\bibnumfmt}[1]{[S#1]}
\renewcommand{\citenumfont}[1]{S#1}
\renewcommand{\thesection}{S\arabic{section}}

\title{Supplementary Online Information - Current-phase relations of few-mode InAs nanowire Josephson junctions}

\maketitle

\setcounter{section}{0}
\setcounter{figure}{0}

\section{Temperature dependence of the CPR}

We measured the CPR of the gated nanowire junction at fixed $V_{BG}$ ($5.55 V$) as a function of temperature (Fig.~\ref{fig:t_dep}). The CPR's amplitude goes to zero between T= 1.02 K and 1.04 K (Fig.~\ref{fig:t_dep} b). The diamagnetic response of the Al ring also goes to zero at T=1.04 K (not shown). At temperatures close to $T_c$, the CPR was sinusoidal, and higher harmonics only contributed to the CPR's shape at lower temperatures (Fig.~\ref{fig:t_dep} a,c). 

The temperature dependencies of CPRs for many nanowires in all three samples show similar behavior: A reduction of both $A_n$ and $a_n$ at high temperatures.

\fig{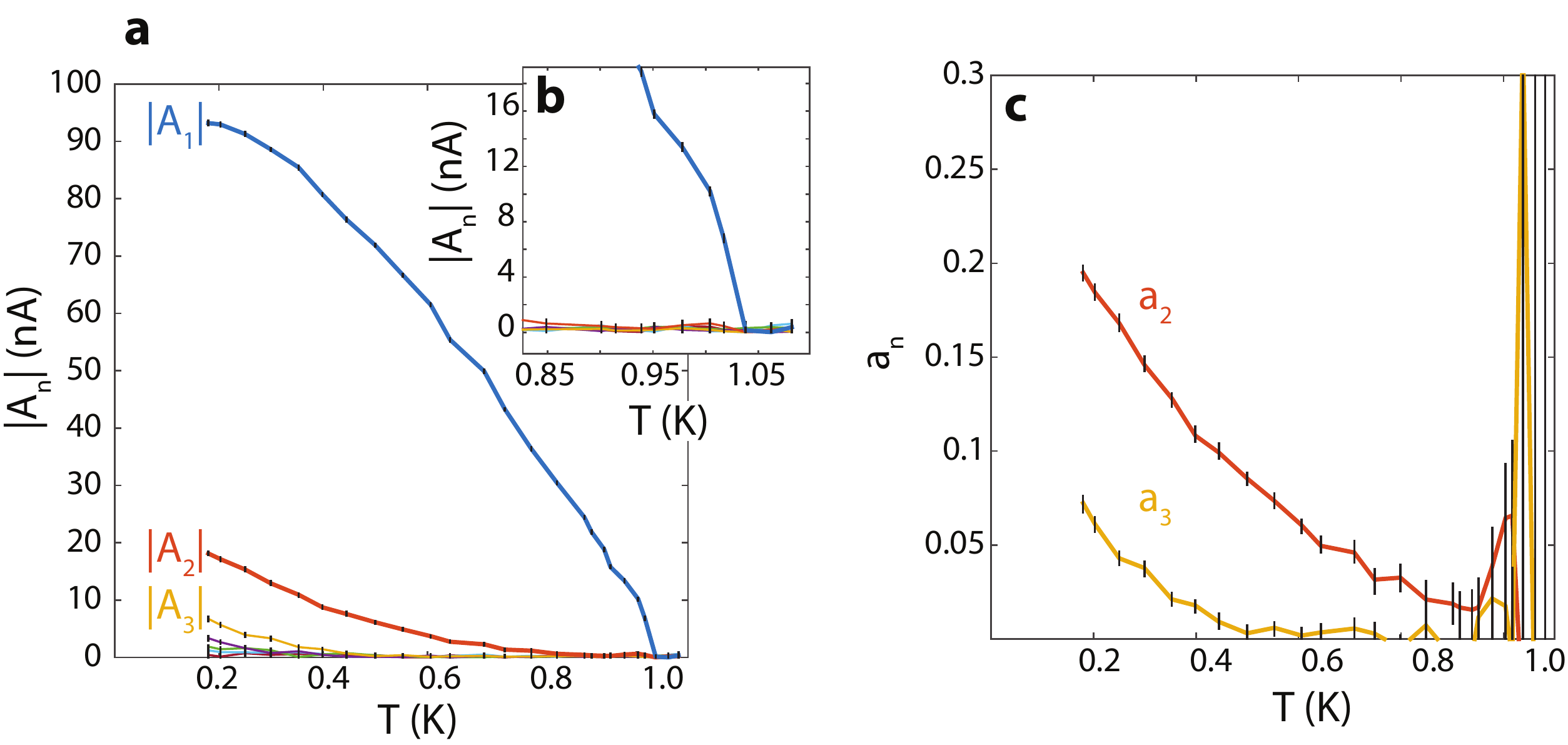}{fig:t_dep}
{Temperature dependence of CPR at fixed $V_{BG}$ }
{(a,b) $A_n$ and (c) $a_n$ vs. temperature at fixed $V_{BG} = 5.55 V$. Error bars are 90$\%$ confidence intervals obtained by bootstrapping 100 times.}{0.9}

The fitted shape parameter $a_2$ of the CPR vs. $V_{BG}$ showed reduced fluctuations at higher temperatures (Fig.~\ref{fig:fluctuations_vs_T}). The average value of $a_2$ at high gate voltages was also reduced with elevated temperature. 

\fig{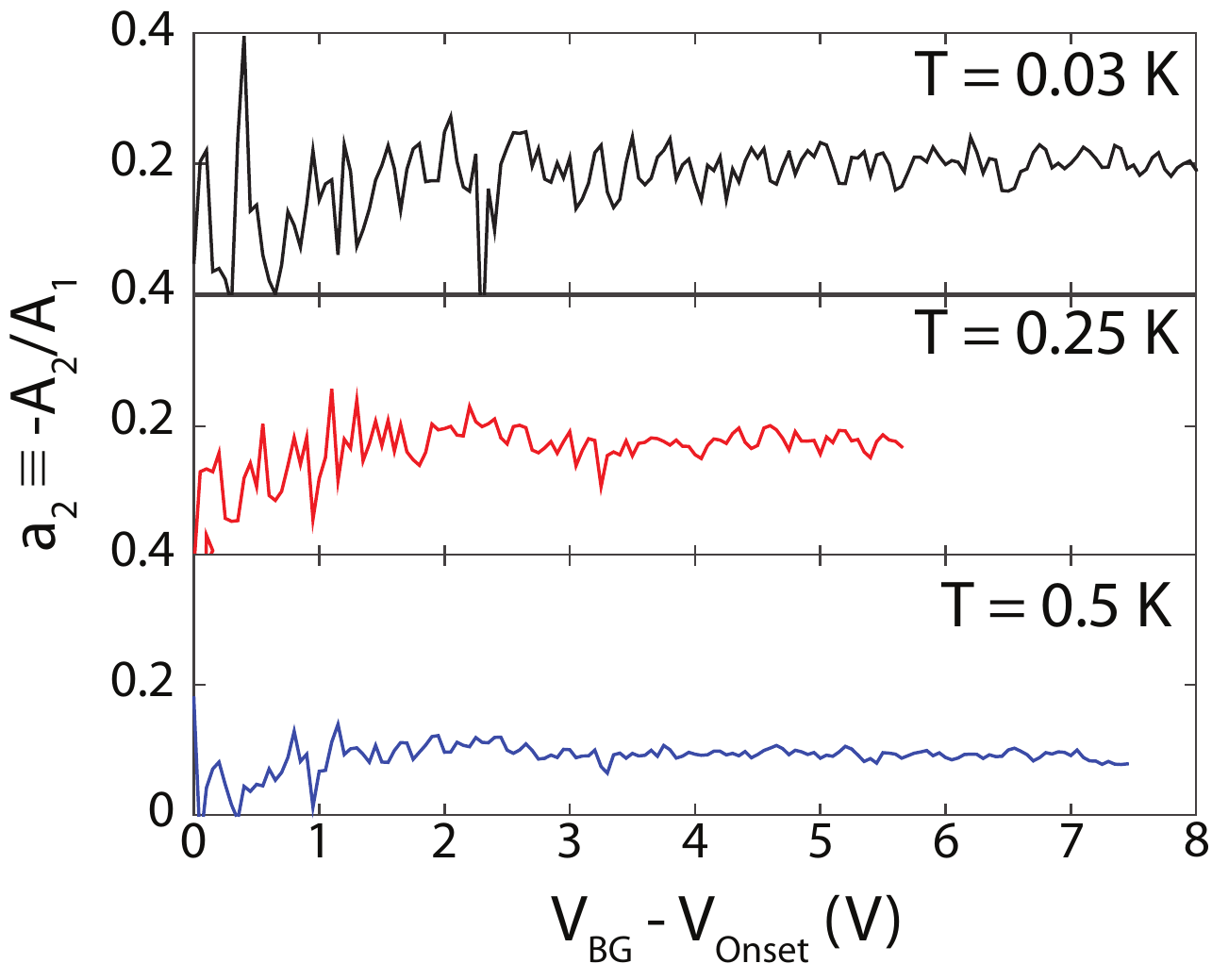}{fig:fluctuations_vs_T}
{Gate voltage dependencies of the forward-skewness at elevated temperatures.}
{The shape of the CPR (as captured by $a_2$) vs. $V_{BG}$ - $V_{Onset}$. $V_{Onset}$ is the gate voltage at which the first non-zero CPR was observed. $V_{Onset}$ = -2.75 V, -1.8 V, and -1.9 V for 30 mK, 250 mK, and 500 mK respectively.}{0.7}

\section{Fitting the most forward-skewed mode}

We fit the CPR shown in Fig.~3c to the expression:

\begin{equation}\label{eq:s1}
I(\phi) = (1+\epsilon)\frac{e \Delta_0(T)}{2 \hbar} \frac{\tau sin(\phi-\phi_0)}{[1-\tau sin^2((\phi-\phi_0)/2)]^{1/2}} tanh(\frac{\Delta_0(T)}{2 k_B T} [1-\tau sin^2((\phi-\phi_0)/2)]^{1/2}), 
\end{equation}

We allowed $T$ to fit due to possible electron heating effects which will not be detected by our thermometry. In previous measurements in our scanning SQUID system in a dilution refrigerator, the electron temperature of devices in a similar geometry was estimated to be 100 mK \citesupp{bluhm}. A fit with fixed $T$ = 0.03 K and $\epsilon = 0$ resulted in large residual structure (Fig.~\ref{fig:fit_perfect} a,b), while a fit with only $T$ fixed gave better results (Fig.~\ref{fig:fit_perfect} c,d), although with a higher error than the fit presented in the main text. 

We plotted $\xi^2 \equiv \sum_{k=1}^N (I_k - I^{fit}_k)^2/N $ vs. $\tau$ and $T_{fit}$ to determine the behavior of errors of our fit (Fig.~\ref{fig:fit_perfect} e). We found that the global minimum is at $T$ = 0.13 K and $\tau = 1.0$, as stated in the main text. Doubling of $\xi^2$ takes place on the intervals $\tau = [0.97, 1.00]$ and $T_{fit} = [0.03,0.13]$, confirming that we have observed a CPR which is consistent with a single, perfectly transmitting Andreev bound state.

\fig{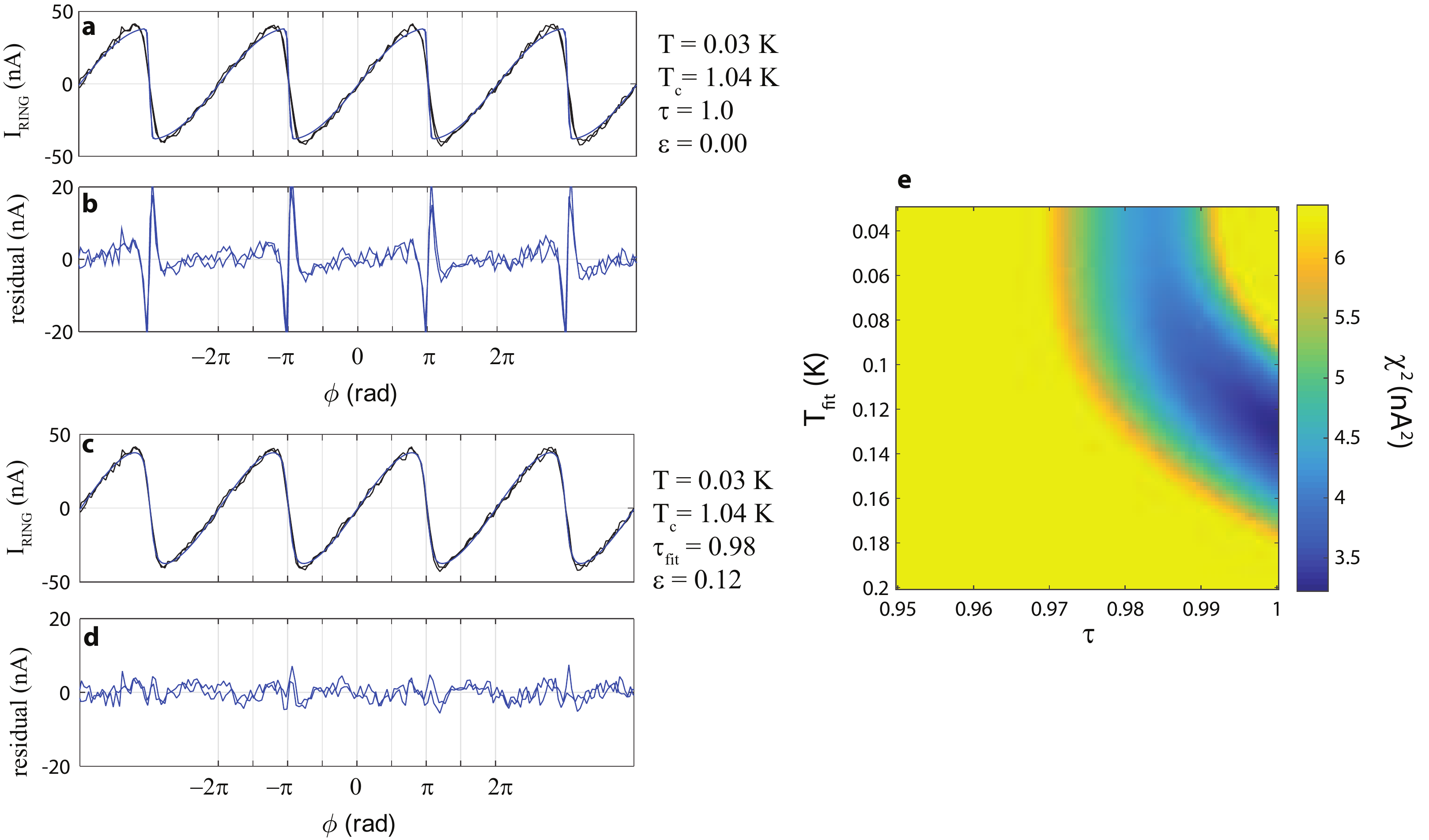}{fig:fit_perfect}
{Results of fits to the most forward-skewed CPR }
{We fit the CPR presented in Fig.~3c to a single-mode short junction expression (Eq.~\ref{eq:s1}). (a,b) Fit and resulting residual with fixed $T$ = 0.03 K and no amplitude scaling factor. This fit, which has obvious residual structure, results in a fitted $\tau = 1.0$. (c,d) Fit and resulting residual with fixed $T$ = 0.03 K and a free amplitude scaling factor ($\epsilon$). This fit, which has much smaller residuals than (b) and results in a fitted $\tau = 0.98$. A full plot of $\chi^2 \equiv \sum_{k=1}^N (I_k - I^{fit}_k)^2/N $ vs. $\tau$ and $T_{fit}$. We allowed $\epsilon$ and $\phi_0$ to vary for each pixel. The colorscale is saturated at two times the observed minimum in $\chi^2$. }{1.0}

\section{Phase shifts in CPRs in Fig.~5}
We encountered phase shifts in our measurements of the CPR, all of which we believe to be instrumental or trivial in nature. For measurements on sample B (Fig.~5), phase shifts between higher harmonics on the same sweep were observed. To account for this, for Fig.~5 we fit to the expression:

\begin{equation}\label{eq:s2}
I(\phi) = \begin{cases} \sum_{n=1}^N A_n sin (n (\phi + \phi^{FW}_n)) &\mbox{if }d\phi/dt > 0 \\ 
\sum_{n=1}^N A_n sin (n (\phi + \phi^{BW}_n)) &\mbox{if } d\phi/dt < 0 \\ 
\end{cases}, 
\end{equation}

which allows for phase shifts between harmonics. Fits to Eq.~\ref{eq:s2} yielded better fits for the CPRs in Fig.~5 (Fig.~\ref{fig:free_phase_fit} c,d). The shape of the CPR in the most forward-skewed CPRs measured on sample B was different between forward and backward sweeps, owing to this effect (Fig.~\ref{fig:free_phase_fit} a,b). The changes in phase for higher harmonics were opposite signs for forward and backward sweeps, indicating that filtering of the signal was to blame. 

Fits without phases (Eq.~3 in main text) were used for Fig.~2-4. Fits to Eq.~\ref{eq:s2} did not change the measured harmonics $A_n$ by more than couple percent even for the highest harmonics, no out of phase component was observed in the FFT (see Fig.~1e), and forward and backward sweeps did not show different shapes. This was in part due to careful tuning of the the PI controller in measurements of sample C, which were not performed for the data taken in Fig.~5. 

\fig{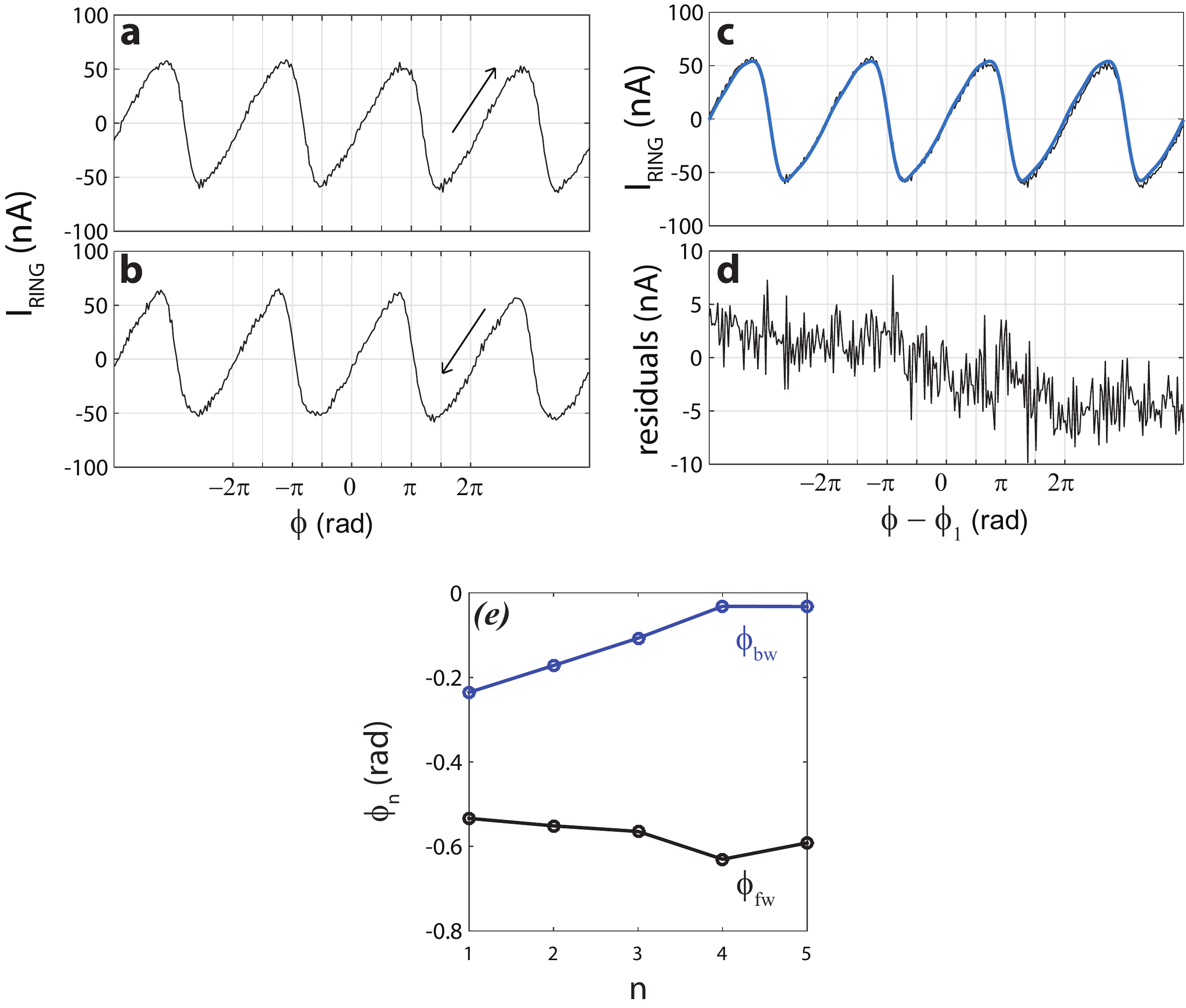}{fig:free_phase_fit}
{Different phases between harmonics are required to correctly fit some CPRs}
{(a,b) Forward and backwards sweep of the same CPR from an L=50 nm ring on sample B. The positive and negative peaks have different shapes, and the shape depends on sweep direction, indicating phase shifts which are different depending on the sweep direction. (c,d) Example of a free phase fit (Eq. \ref{eq:s2}, blue line) to the forward sweep and the residuals. (e) The fitted phases of the harmonics vs. harmonic number n. }{0.9}

\section{Hysteretic behavior}

Many rings were in the hysteretic regime at low temperatures (e.g., Fig.~\ref{fig:hysteretic}). For the majority of the rings where hysteresis was observed, obvious underetching of the nanowire resulted in Al bridging the would-be junction (Fig.~\ref{fig:hysteretic} inset). 

The open shape of the measured response of the ring is due to the presence of multiple local minima in the Josephson energy, which arise only in the when $\beta \equiv 2\pi L_{self} I_c / \Phi_0 > 1$ \citesupp{jackel2}. The height of the individual jumps is given by $I_{jump} = \Phi_0 / 2 L_{self} $, while total maximum current is just given by the critical current $I_C$. The height of the jumps was similar for many rings and consistent with the calculated self inductance of 13.9 pH. The overall height of the hysteretic pattern varied from ring to ring, indicating that the critical current of the junctions varied.

\fig{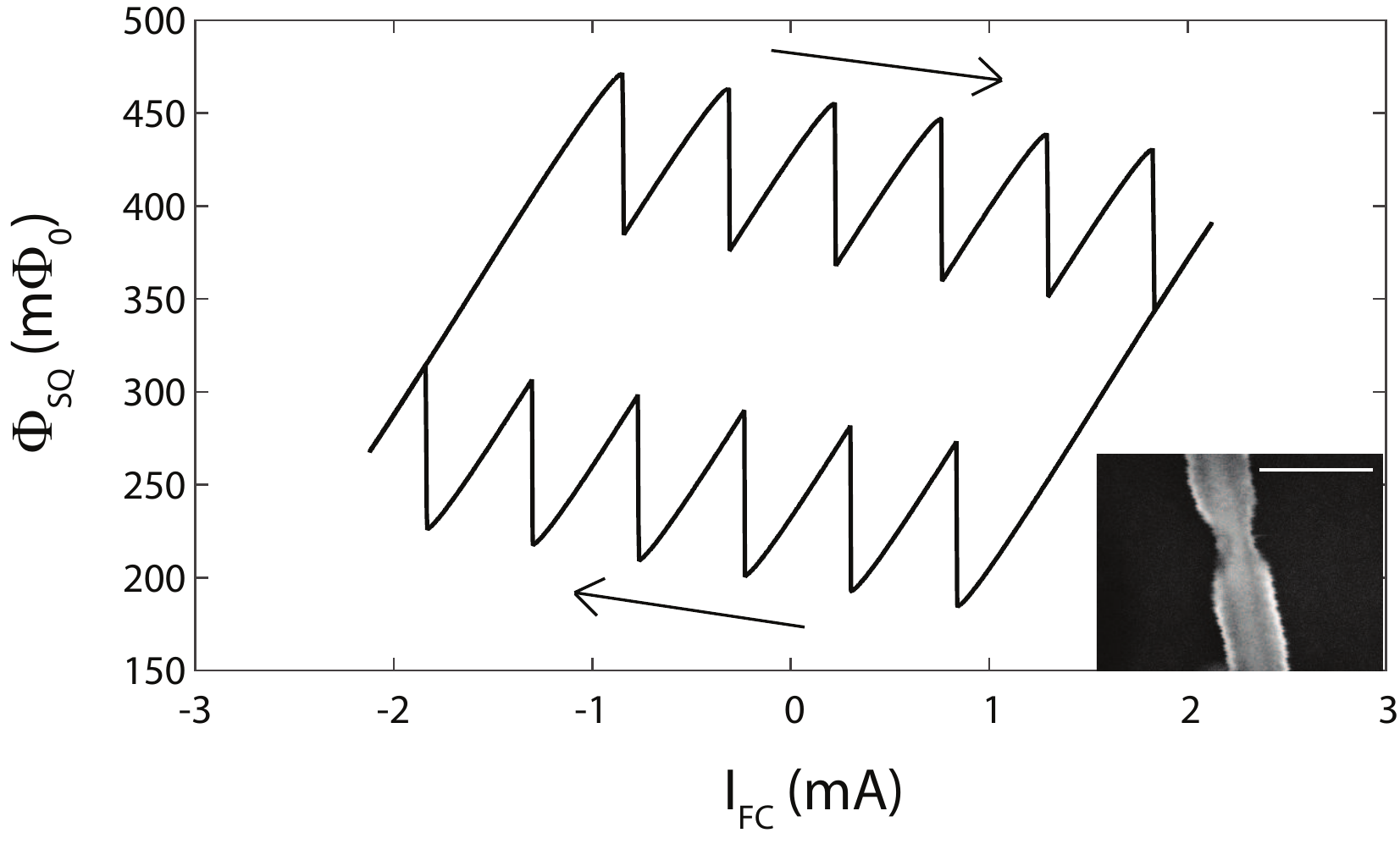}{fig:hysteretic}
{Example of a hysteretic ring}
{A typical measurement of the measured flux through the SQUID ($\Phi_{SQ}$) as function of current through the field coil ($I_{FC}$) on a hysteretic ring on sample C. Arrows indicate the sweep direction of the field coil current. (Inset) SEM image of the junction measured in the main figure, showing under-etching of the Al which resulted in a very high critical current and hysteretic behavior. Scale bar is 200 nm}{0.9}

\section{Simulations}

We performed two types of simulations to elucidate the behavior we've observed in the CPR. The first, a simulation based on Ref. \citesupp{FurusakiPRB2}, involves simulating an InAs 2DEG with similar length and width to the wire we studied. This simulation is useful because it does not rely on the short junction approximation to calculate the CPR, and it allows us to introduce interfacial barriers which depend on the Fermi velocity mismatch between the superconductor and the normal region. Because the geometry is only loosely related to the experimental geometry, the subband structure, relative spacing of resonances, and exact dependence of the mismatch on gate voltage are not well simulated, however, it does give us some useful intuition.

We made minor modifications to the calculation performed in section V of Ref. \citesupp{FurusakiPRB2} to calculate the CPR vs. electron density in the nanowire. In particular, we calculated $Z_i$, the effective interface transparency, for each subband $i$, which is given by $Z_i=(1-r_i)^2/(4r_i)$, where $r_i \equiv v_F^S / v_{F,i}^N$, the ratio of the Fermi velocity in the superconductor and the ith subband of the normal material. We fitted the calculated CPR to extract $A_n$ and $a_n$ in a similar way to real data. 

The low density, single subband behavior exhibits peaks that match what we observed experimentally close to depletion of the nanowire (Fig.~\ref{fig:furusaki} a,b). In particular, the observed shape of the peaks in $A_n$ and $a_n$ are asymmetric and cusp-like. Here, the asymmetry arises due to the density-dependent barrier, which decreases at higher densities due to a smaller Fermi velocity mismatch. The height of $a_2$ for all peaks except the first peak are $\approx 0.4$, consistent with our the maximum allowed forward-skew in short junction theory and the observed peak in Fig.~3. 

At much higher densities, the resonant peaks begin to look more like oscillations, due to a decrease in the Fermi velocity mismatch (Fig.~\ref{fig:furusaki} c,d). The behavior of other subbands are exactly the same as in (Fig.~\ref{fig:furusaki} c,d), but offset by the energy onset of each subband. The result is the behavior observed in (Fig.~\ref{fig:furusaki} e,f), which shows peak-like behavior at low densities, but more fluctuation-like behavior when multiple subbands are occupied. 

\fig{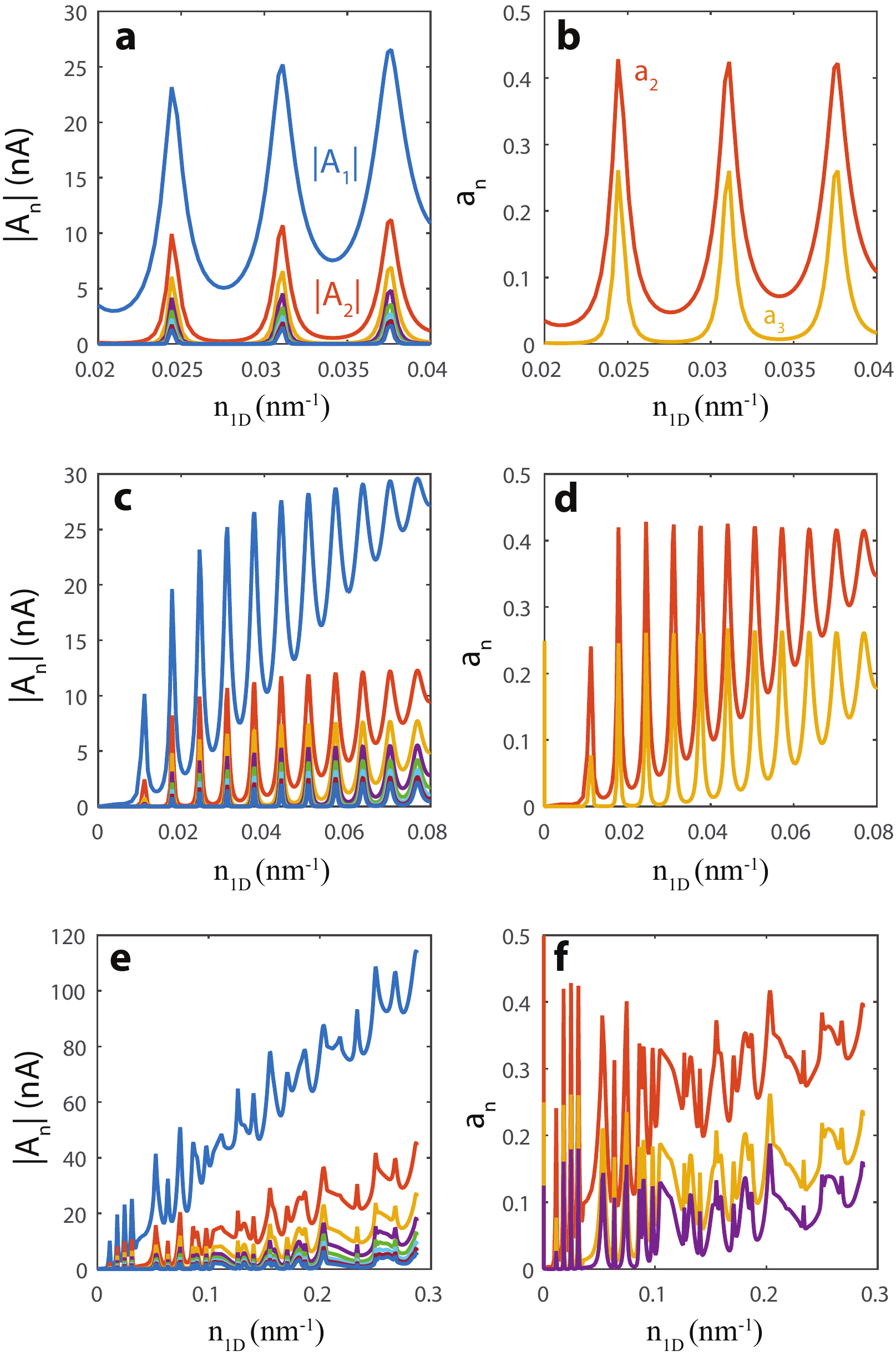}{fig:furusaki}
{Simulations following Ref. S30}
{Simulations of the CPR for an InAs 2DEG with parameters: $w_{InAs} = 50$ nm, $w_{SC} = 110$ nm, $m^* = 0.023 m_e$, $L = 150$ nm, $v_F^S = 1.5 \times 10^6$ m/s . An interfacial barrier which scales with the mismatch in Fermi velocities between the superconductor and semiconductor resulted in Fabry-Perot like oscillations in the Fourier amplitudes and shape parameters (See Text). The fitted CPR parameters $A_n$ and $a_n$ for a single subband (a-d) and many subbands (e,f).}{0.7}

We performed effective tight-binding simulations using the Kwant Python package \citesupp{kwant2}, using a more realistic nanowire geometry (Fig.~\ref{fig:kwant} a). In the real nanowire, the thickness of the nanowire is increased due to epitaxial Al, however in our simulation everything is InAs. Wave function mismatch should occur in both, due to different spatial distribution of the wave functions and differences in chemical potential. In the physical nanowire, Al should lead to strong band bending which causes the wave function probability to exist mostly near the Al/InAs interface. In the tight binding model, the wave function mismatch is simulated by a change in the diameter of the nanowire and an applied offset to the chemical potential in the thicker regions of the nanowire.

We solved the following Hamiltonian on a square lattice with a lattice spacing $a = 5nm$:
\begin{equation}\label{eq:s3}
H= \sum_{i,j} (6t - \mu_{i,j}) \ket{i,j}\bra{i,j} - t (\ket{i+1,j}\bra{i,j}+ \ket{i,j}\bra{i+1,j}+\ket{i,j+1}\bra{i,j}+\ket{i,j}\bra{i,j+1}), 
\end{equation} where $t=\hbar^2 / 2 m^* a^2$ and $\mu$ was varied as a function of space. In the thick part of the nanowire, the chemical potential was held fixed at $\mu = 800$ meV. In the gated region $\mu$ was varied to tune the density of the nanowire. We applied a gradient to $\mu$ along the z direction to mimic the effect of a close bottom gate, which acts to split up normally degenerate subbands. Defining z=0 at the center of the nanowire, we set $\mu$ inside the nanowire to $\mu_{etched}+\mu_{etched} z/4r$, where r is the radius of the nanowire. 

Using Kwant's solver, we obtained the normal state S-matrix and calculated the transmission eigenvalues at each $\mu_{core}$. We then used the short junction equation (Eq.~\ref{eq:s1}) to calculate the CPR and fitted it to extract $A_n$ and $a_n$ (Fig.~\ref{fig:kwant} b-e). The fitted CPR parameters display qualitative behavior very similar to the measured CPR (Fig.~2), specifically resonant peaks at low density, a very forward-skewed mode at low densities, and fluctuation-like behavior at higher densities. Without added onsite disorder, the simulated forward-skewness observed at high densities are very close to the experimental result ($a_{2,exp} \approx 0.21$ and $a_{2,sim} \approx 0.2$). Random onsite disorder leads to an increase in the frequency of fluctuations, presumably due to extra induced resonant behavior, as well as a reduction of the forward-skewness at high densities, as is expected for a ballistic to diffusive transition.

The observed gate voltage dependence of the CPR and the presence of a highly forward-skewed mode are reproduced in both WKB and tight-binding simulations in the presence of a Fermi velocity or wave function mismatch. The addition of disorder to our model is not required to reproduce the qualitative behavior of fluctuations we observed.

\fig{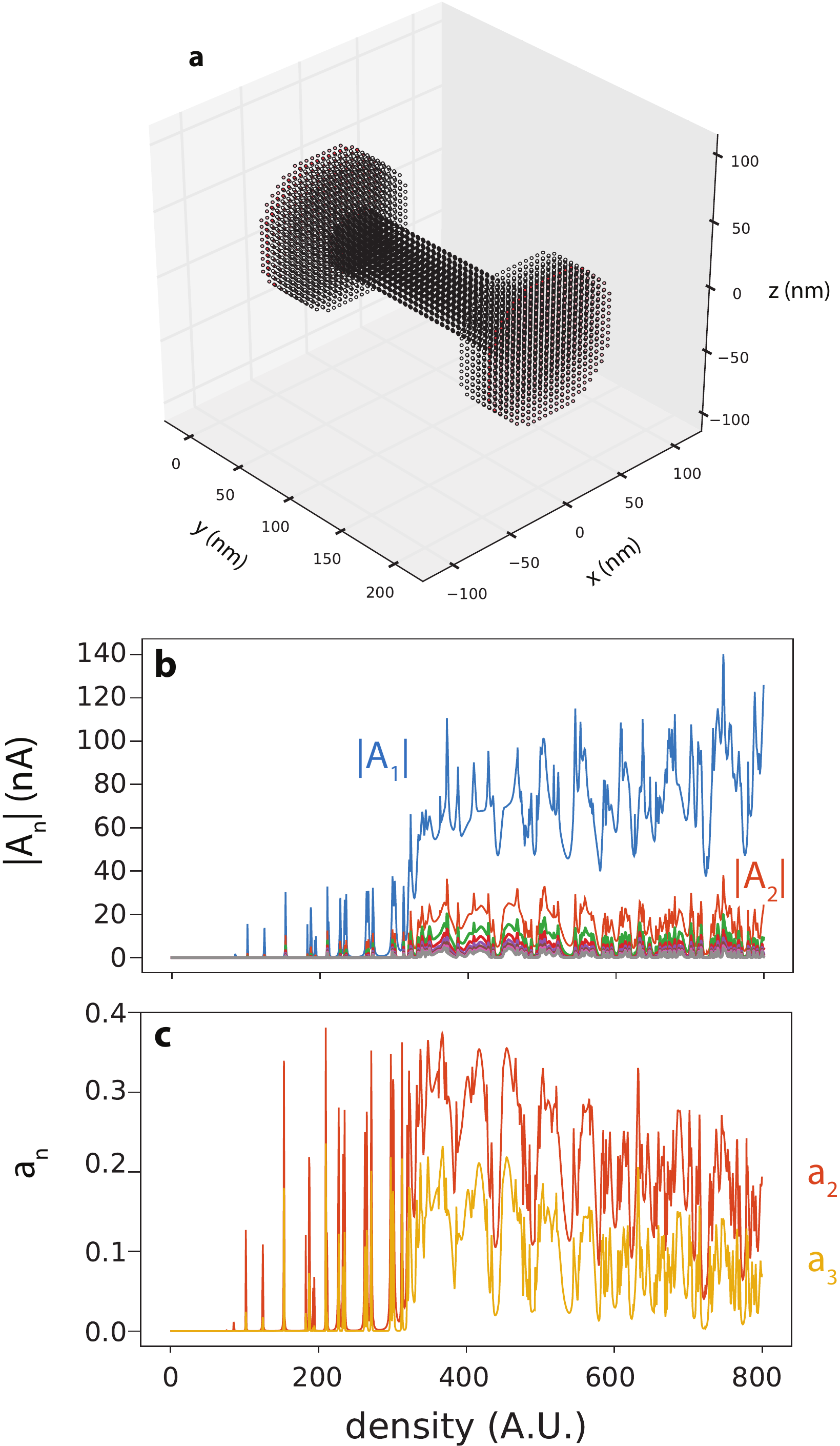}{fig:kwant}
{Effective tight-binding simulations of the CPR of an InAs nanowire}
{Effective tight-binding simulation of the CPR for an InAs nanowire with parameters: $w_{InAs} = 50$ nm, $w_{SC} = 110$ nm, $m^* = 0.023 m_e$, $L = 150$ nm, $r=25$ nm to $55$ nm. (a) A plot of the effective tight binding geometry, the chemical potential in the  dark region was modulated while the light region was fixed at $\mu = 400$ meV. (b,c) Fourier amplitude $A_n$ (b) and shape parameter, $a_n$ (c) vs. total density of states in the nanowire.}{0.68}

\FloatBarrier
\section{Amplitude vs. Skew}

\begin{figure}[htb]
	\centerline{\includegraphics[width=0.5\textwidth]{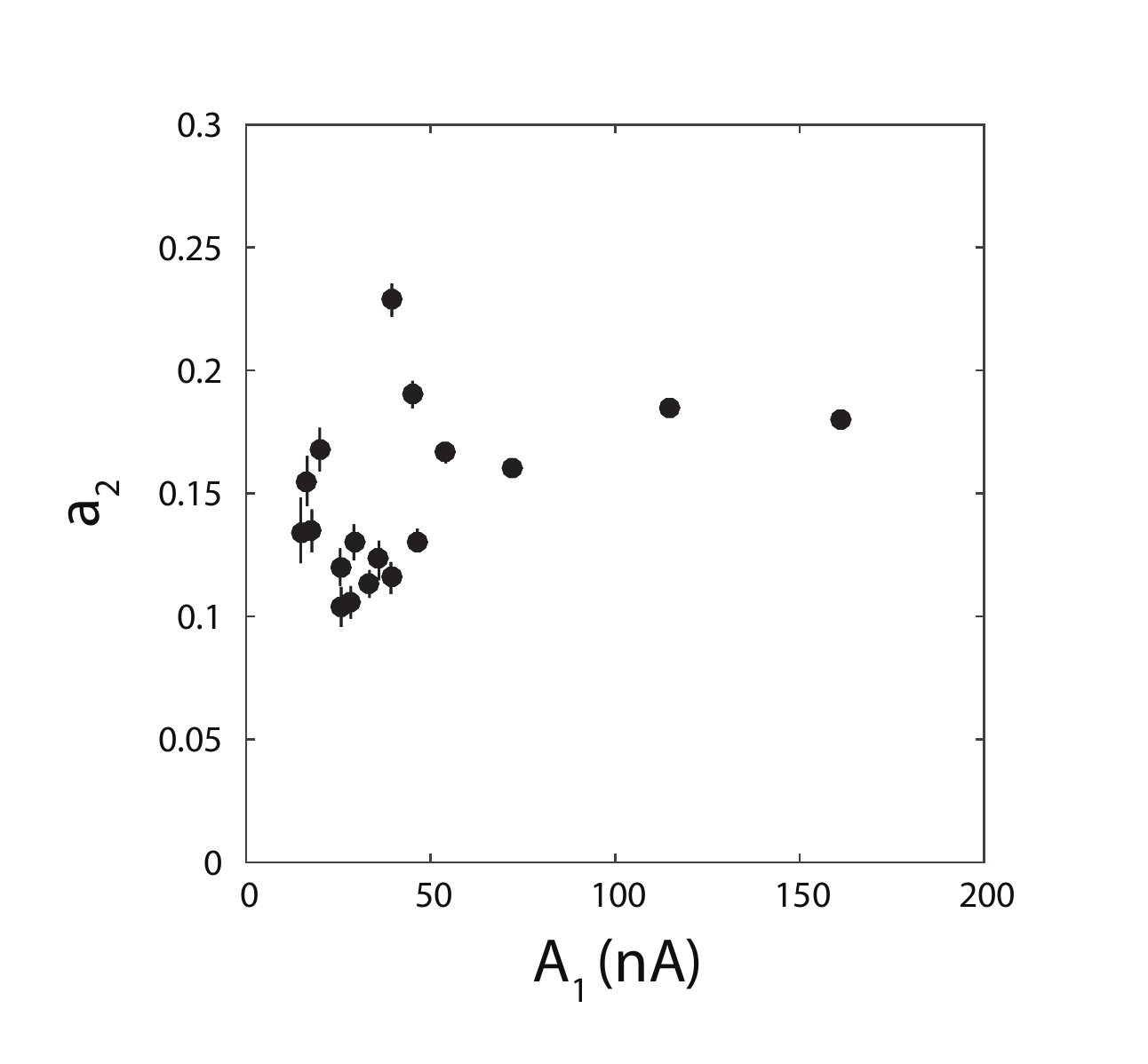}}
\caption[Shape parameter ($a_2$) vs. CPR amplitude ($A_1$)]{\label{fig:A1_vs_a2} \textbf{Shape parameter ($a_2$) vs. CPR amplitude ($A_1$)} Fitted $a_2$ vs. $A_1$ for many rings at $V_{BG}$ = 0 V and $T  <$ 50 mK. The CPRs fitted here are the same CPRs presented in Fig.~5.}
\end{figure}

\FloatBarrier
\bibliographysupp{refs}

\end{document}